# Surface acoustic microwave photonic filters on etchless lithium niobate integrated platform

Yue Yu and Xiankai Sun[*]

*Department of Electronic Engineering, The Chinese University of Hong Kong, Shatin, New Territories, Hong Kong*

[*]*Corresponding author:* xksun@cuhk.edu.hk

**Lithium niobate on insulator emerges as a promising platform for integrated microwave photonics because of its capability of ultralow-loss guidance and high-efficiency modulation of light. Chip-level integration of microwave filters is important for signal processing in the 5G/6G wireless communication. Here, we employed the principle of bound states in the continuum for low-loss waveguiding on an etchless lithium niobate integrated platform, and realized high-performance microwave photonic filters thereon. These microwave photonic filters consist of a high-quality photonic microcavity modulated by piezoelectrically excited surface acoustic waves. Acoustic time delays from 21 to 106 ns and passbands with bandwidth as narrow as 0.89 MHz were achieved in the fabricated filters operating at gigahertz frequencies. Our demonstration may open up new applications on the lithium niobate integrated platform, such as optical communication, signal processing, and beam steering.**

## 1. Introduction

Integrated microwave photonics by combining photonic integrated circuit technology with microwave photonics (MWP) has shown its advantages in enhanced functionalities and robustness as well as reduced size, weight, cost, and power consumption. Microwave filtering is one of the most important functionalities in integrated MWP for signal processing in the 5G/6G wireless communication[1,2], which separates signals of interest from the noise background and mitigates unwanted interference[3]. A common way for implementing integrated MWP filters is constructing multitap delay lines on a chip [4,5], where the input signal is discretely sampled, delayed, and weighted before being summed up. For applications that require narrow linewidths and/or free spectral ranges (FSRs), long delays become necessary, leading to large footprints and high losses. Hypersonic acoustic waves with significantly lower velocity than optical waves can be adopted for overcoming this difficulty and obtaining the required long delays within a small footprint. One way of harnessing the acoustic waves for high-resolution filtering relies on the stimulated Brillouin scattering between guided light and sound waves, usually in chalcogenide glasses[6,7] and suspended silicon membranes[8-10]. Another type of sound waves, surface acoustic waves (SAWs) propagate at the surface of a material within a depth on the order of acoustic wavelength. SAWs have very high energy confinement and can attain large overlap with the optical mode of a planar waveguide on integrated platform, and thus have been widely applied in the areas of quantum communication[11], sensing[12,13], and signal processing[14-16]. Therefore, SAWs can be an excellent candidate for obtaining the long delays in integrated MWP filters.

To date, most of the integrated MWP components are realized on mature material platforms including indium phosphide[17,18], silicon on insulator[15,16,19], and silicon nitride on insulator[20,21]. Lithium niobate on insulator (LNOI) has recently emerged as a promising platform for integrated photonics due to its excellent material properties including strong electro-optic, acousto-optic, and nonlinear optical effects, as well as its capability for strong optical confinement and high photonic integration density[22,23]. Recently, a fundamentally new photonic architecture was demonstrated by patterning a fabrication-friendly optically transparent material on single-crystal thin-film substrate without the need for etching of the substrate[24,25]. This architecture has enabled various photonic functionalities, including 2D-material integration[26,27], high-dimensional communication[28], acousto-optic modulation[29,30], and second-harmonic generation[31], with performance comparable with those on etched LNOI platform. Here, by monolithically integrating an interdigital transducer (IDT) with a multitap photonic microcavity, we demonstrated high-resolution MWP filters on an etchless LNOI integrated platform for signal processing. The demonstrated filters can be scaled for operation in a wide frequency range from megahertz to tens of gigahertz, which will contribute to the development of high-performance MWP system on the lithium niobate integrated platform.

## 2. Results and Discussion

Figure 1(a) illustrates the surface acoustic MWP filter fabricated on a $z$-cut LNOI substrate with an etchless process. We patterned a multitap photonic microcavity, where the taps consist of multiple equidistantly located straight waveguides joined by 180°-bends, in a polymer (ZEP520A) on an LNOI substrate. An IDT was fabricated in the vicinity of an outermost tap for exciting the SAWs. To obtain maximal acousto-optic modulation, the IDT is oriented along the $x$ direction by considering the anisotropic photoelastic coefficients in Ref. 32. Figure 1(b) shows the cross section of the acousto-optic interaction region, where the thicknesses of the IDT gold electrodes, the lithium niobate layer, the polymer atop, and the silicon oxide underneath are 80 nm, 300 nm, 400 nm, and 2 μm, respectively. $w$, $d$, $p$, and $t$ represent the width of the polymer waveguide, the distance between adjacent taps, the period of the IDT, and the width of the IDT fingers, respectively. Figure 1(c) shows a top view of the photonic microcavity and IDT, where $l$ is the aperture of the IDT and is also the length of the straight waveguide in a tap. An input microwave signal with frequency $f$ was applied to the IDT to generate SAWs of the same frequency. When $pf$ matches the velocity $v$ of an SAW, the SAW can be excited and propagate away from the IDT (See Supplementary Sec. 2), which induces a phase modulation to light as it crosses a tap due to the acousto-optic effect. Considering that the typical velocity of SAWs is in the range of several thousand meters per second, an acoustic propagation length of hundreds of micrometers between adjacent taps induces a time delay of hundreds of nanoseconds. The phase modulation accumulated from $N$ taps is converted to an intensity modulation at the output of the photonic microcavity. The impulse response of the MWP filter can be described by

$$h_R(t) = K \sum_{m=0}^{N-1} \exp\left(-\frac{\alpha}{2} m v T\right) \delta(t - mT),$$

where $K$ is a prefactor that includes the efficiency of SAW excitation, acousto-optic modulation, and photodetection, $\alpha$ is the propagation loss coefficient of the SAW, and $T$ is the unit acoustic delay equal to $d/v$ (See Supplementary Sec. 1). The properties of the MWP filter are determined by the geometry of the photonic microcavity and the IDT. For example, a smaller FSR in the transfer function can be obtained by increasing the spacing between adjacent taps, a narrower bandwidth can be obtained by increasing the number of taps, the number of output passbands can be reduced by increasing the periods of the IDT (also see Supplementary Sec. 1). The surface acoustic MWP filters with design flexibility can facilitate microwave signal processing on the lithium niobate integrated photonic platform.

A major motivation for the MWP development is the low-loss transmission over long distances. The optical loss can have considerable effects on the system's performance, because it transforms into the microwave loss quadratically due to the square law in optical-to-electrical conversion by photodetectors[33]. Therefore, low-loss photonic integrated circuits are the foundation for constructing high-performance MWP filters on a chip. For a waveguide structure made of a low-refractive-index material on a high-reflective-index substrate as shown in Fig. 1(b), the TM-polarized bound mode lies in the TE-polarized continuous spectrum. Therefore, the TM bound mode can interact with the TE continuum, resulting in energy dissipation of the former. However, one can engineer the waveguide geometry to turn the TM bound mode into a bound state in the continuum (BIC) with completely eliminated energy dissipation[24]. To obtain a high-quality photonic microcavity with multiple taps as shown in Fig. 1(a), one should minimize the waveguide propagation loss in all the straight and bent sections. The upper panel of Fig. 2(a) shows that the propagation loss of the fundamental TM mode in a straight waveguide depends on the waveguide width $w$. A zero propagation loss can be achieved at $w = 1.74$ μm at the wavelength of 1510 nm. On the other hand, the propagation loss of the fundamental TM mode in a bent waveguide depends on both the waveguide width $w$ and the bend radius $R$. The lower panel of Fig. 2(a) plots the simulated propagation loss of the TM mode in a bent waveguide as a function of the bend radius $R$ with $w = 1.74$ μm. When $R$ is larger than 60 μm, a zero propagation loss can be obtained at periodical $R$ values ($R = 78, 92, 106$ μm …). Figure 2(b) shows the modal profiles of the fundamental TM mode in a straight waveguide (upper) and a bent waveguide (lower) at the BIC and off-BIC points marked in Fig. 2(a). For the TM mode at the off-BIC point, there is obvious energy dissipation to the TE continuum. By contrast, the TM mode at the BIC point exhibits perfectly localized field, resulting in lossless propagation of photons in the continuum.

Based on the simulated results, we designed the multitap photonic microcavity by satisfying the BIC conditions for both straight and bent waveguides ($w = 1.74$ μm, $R = 134$ μm). We also designed the nearby SAW IDT with $p = 2w = 3.48$ μm unless otherwise noted, to achieve maximal acousto-optic modulation efficiency. We fabricated the devices with an etchless process on a $z$-cut LNOI substrate. First, the IDT was fabricated with a liftoff process, which involved pattern definition by high-resolution electron-beam lithography and the subsequent gold deposition by electron-beam evaporation. Then, the polymer waveguides and microcavity were fabricated with a second step of electron-beam lithography. Note that in this step the polymer on the SAWs' propagation path was also removed to avoid its attenuation effects on the SAWs. Figure 3(a) shows a top-view optical microscope image of a fabricated 4-tap device, and Fig. 3(b) is a close-up view of its acousto-optic modulation region. We first measured the optical transmission of the photonic microcavity by coupling light into and out of the device via grating couplers. The input light was provided by a tunable semiconductor laser (Yenista T100S-HP) and the output light was collected by a photodetector (Hewlett Packard 81531A). Figure 3(c) plots the optical transmission spectrum of the device shown in Fig. 3(a). The zoomed-in spectrum in Fig. 3(d) indicates loaded quality factors $Q_L$ of ~$1.2 \times 10^5$ near the wavelength of 1510.0 nm. The measured optical FSR is 0.355 nm, which agrees well with the theoretical value of 0.358 nm obtained by using the cavity circumference of 2820 μm and the simulated group index of 2.253.

Figure 4(a) shows the experimental setup for characterizing microwave filtering of the fabricated devices. Light from the tunable semiconductor laser had its polarization adjusted by a fiber polarization controller before being coupled into the device via the input grating coupler. The laser wavelength was tuned to a maximal slope of the photonic microcavity's transmission spectrum. Meanwhile, a sinusoidal microwave signal from a vector network analyzer (Keysight E5071C) was delivered to the IDT via a microwave probe. The light inside the photonic microcavity experienced multiple times of acousto-optic modulation and was strongly enhanced near a cavity resonant wavelength. The light coupled out of the output grating coupler was split and collected by two photodetectors. A high-sensitivity photodetector (Hewlett Packard 81531A) was used to monitor the cavity transmission for tuning the laser wavelength. A high-speed photodetector (Optilab A1803) was used to covert the SAW-modulated light signals into the electrical domain, which were sent back to the vector network analyzer for obtaining the transfer function.

Figure 4(b) shows the measured $S_{11}$ spectrum, which shows several dips in the measured frequency range. To identify the excited SAW modes, we fabricated a series of devices with IDT finger period $p$ varying from 2.6 to 4.2 μm. Figure 4(c) shows the simulated and measured frequencies of the SAW modes, which agree well with each other. The red dashed line marks the case in Fig. 4(b), where the IDT finger period is $p = 2w = 3.48$ μm, and the two red stars represent the measured frequencies at the two dips [0.78 GHz (①) and 1.60 GHz (②)]. Figure 4(d) shows the cross-sectional displacement field profiles ($x$ component) of the above two SAW modes. Based on the relationship between the IDT finger period and the excited SAW frequency in Fig. 4(c), we obtained the simulated velocity of SAW modes ① and ② as $v_1 = 2714$ m s$^{-1}$ and $v_2 = 5398$ m s$^{-1}$, respectively (see Supplementary Sec. 2).

Next, we characterized the microwave filtering performance of our fabricated devices. The incident microwave drive power was set as −5 dBm for all measurements. Figures 5(a) and 5(b) show the measured normalized $|S_{21}|$ spectra of 4- and 6-tap surface acoustic MWP filters with $d = 2R = 268$ μm at frequencies near 0.78 and 1.60 GHz. These spectra exhibit periodic passbands with an FSR of 10.6 and 16.6 MHz, corresponding to an acoustic delay of 94.3 and 60.2 ns, respectively, over the propagation distance $d$. Additionally, the passbands' bandwidth decreases as the number of the taps increases, achieving the smallest obtained full width at half maximum of 0.89 and 2.9 MHz, respectively. Figures 5(c) and 5(d) plot the normalized impulse response, derived from the measured complex-valued frequency response in Figs. 5(a) and 5(b), respectively. Exponential fits of the decaying amplitude of the impulse response in Figs. 5(c) and 5(d) indicate the propagation loss $\alpha$ of 5.2 and 2.1 mm$^{-1}$ for the SAW modes ① and ②, respectively.

Since the FSR of the passbands depends on both the SAW velocity $v$ and the tap spacing $d$, we fabricated a series of 6-tap surface acoustic MWP filters with different tap spacings $d$, where $R = d/2$ satisfies the BIC condition for bent waveguides as shown in Fig. 2(a). We experimentally achieved delay times from 21 to 106 ns and measured the velocities of the two excited SAW modes as 2842 m s$^{-1}$ and 4904 m s$^{-1}$ (see Supplementary Sec. 2). With these experimental results, we estimated the effective refractive index perturbation magnitude as $\Delta n_{\text{eff}} = 4.7 \times 10^{-6}$ RIU and the acousto-optic modulation depth as $M = 4.75\%$ (see Supplementary Sec. 3). These results can be further improved by optimizing the fabrication process to reduce the optical loss and by optimizing the IDT structure to enhance the SAW excitation efficiency.

## 3. Conclusion

In conclusion, we have experimentally demonstrated surface acoustic MWP filters on an etchless lithium niobate integrated platform. The devices consist of a high-quality multitap photonic microcavity modulated by piezoelectrically excited surface acoustic waves. The devices operated under the principle of photonic bound states in the continuum for low-loss waveguiding and high-quality resonating and achieved high-performance in microwave filtering. We fabricated devices with 4 and 6 taps and realized an acoustic delay from 21 to 106 ns. Compared with other types of MWP filters based on Brillouin scattering and/or in a fiber-optic system, our surface acoustic MWP filters benefit from these competitive advantages: (i) simple fabrication process which does not need etching of the substrate or suspension of delicate membranes or waveguides; (ii) long delay achieved within a small footprint due to the slow velocity of acoustic waves; (iii) robustness against environmental perturbations such as vibrations and temperature gradients; (iv) easy extension of the microwave filtering concept to any other substrates and/or frequencies. With the superior properties of lithium niobate, the demonstrated surface acoustic MWP filters pave the way for a fully integrated MWP system on an LNOI chip, including light sources, amplifiers, modulators, signal processors, and photodetectors, with significantly enhanced compactness and system stability.

**Acknowledgment**

This work was supported by Research Grants Council of Hong Kong (No. 14206318, 14208421), and by Strategic Partnership Award for Research Collaboration offered by The Chinese University of Hong Kong. The authors would like to thank Mr. Piao Deng for his assistance on theoretical analysis.

**Competing interests**

The authors declare no competing interests.

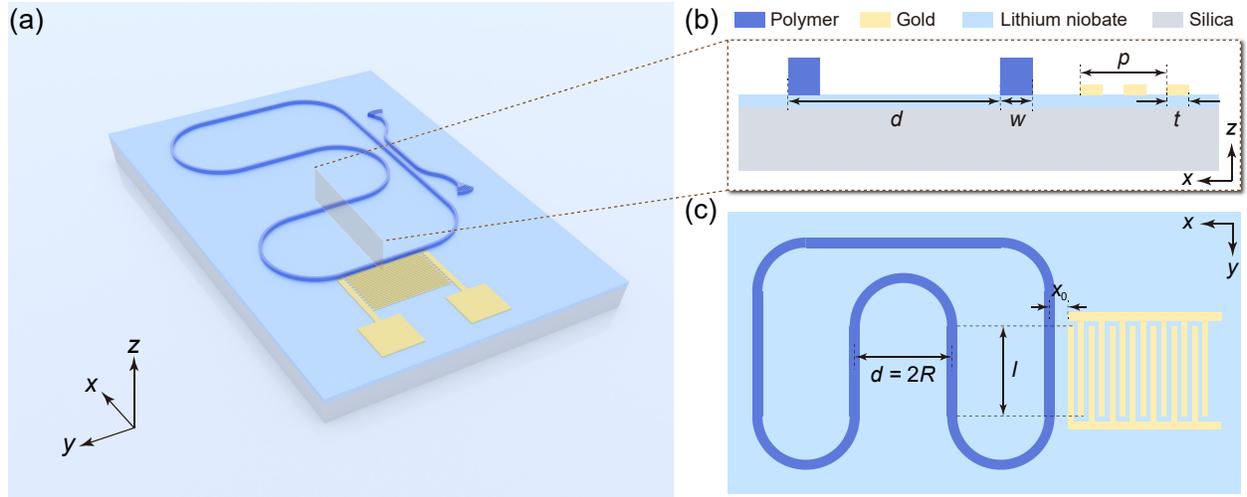

**Fig. 1.** (a) Illustration of the surface acoustic microwave photonic filter on etchless LNOI platform. (b) Cross-sectional view of the acousto-optic modulation region near the IDT. The dark blue part denotes the polymer waveguide, the yellow part denotes the IDT made of gold, the light blue part denotes the lithium niobate layer, and the gray part denotes the bottom silicon oxide (the insulator). $w$, $d$, $p$, and $t$ represent the width of the polymer waveguide, the distance between adjacent taps, the period of the IDT, and the width of the IDT fingers, respectively. (c) Top view of the photonic microcavity and IDT, where $x_0$, $d$, and $l$ denote the distance between the IDT and the first tap, the distance between adjacent taps, and the aperture of the IDT, respectively.

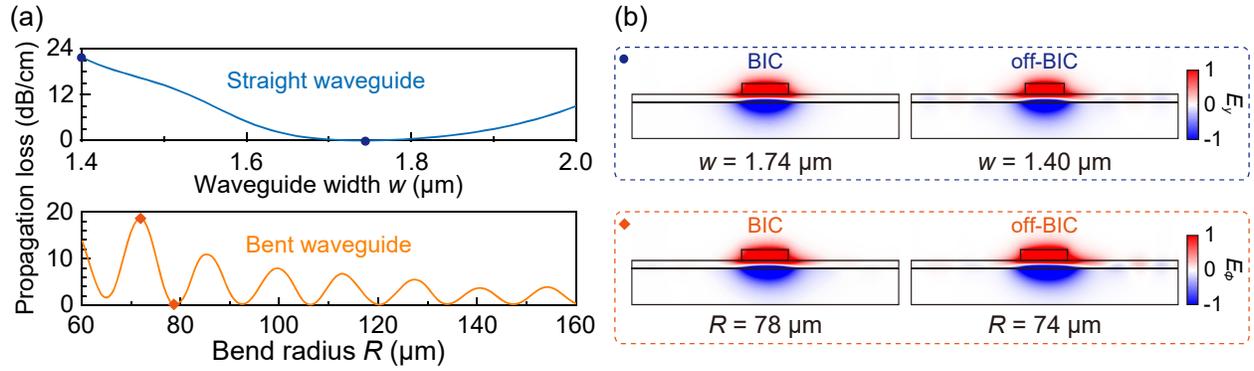

**Fig. 2.** (a) (Upper) Calculated propagation loss of a straight waveguide as a function of waveguide width $w$. (Lower) Calculated propagation loss of a bent waveguide as a function of bend radius $R$ with fixed $w = 1.74$ μm. In both cases, the simulation was performed at the wavelength of 1510 nm. (b) Modal profiles of the TM bound mode in a straight waveguide (upper) and in a bent waveguide (lower) at the BIC and off-BIC points marked in (a). $E_y$ ($E_\varphi$) denotes the electric field component along the propagation direction of the straight (bent) waveguide.

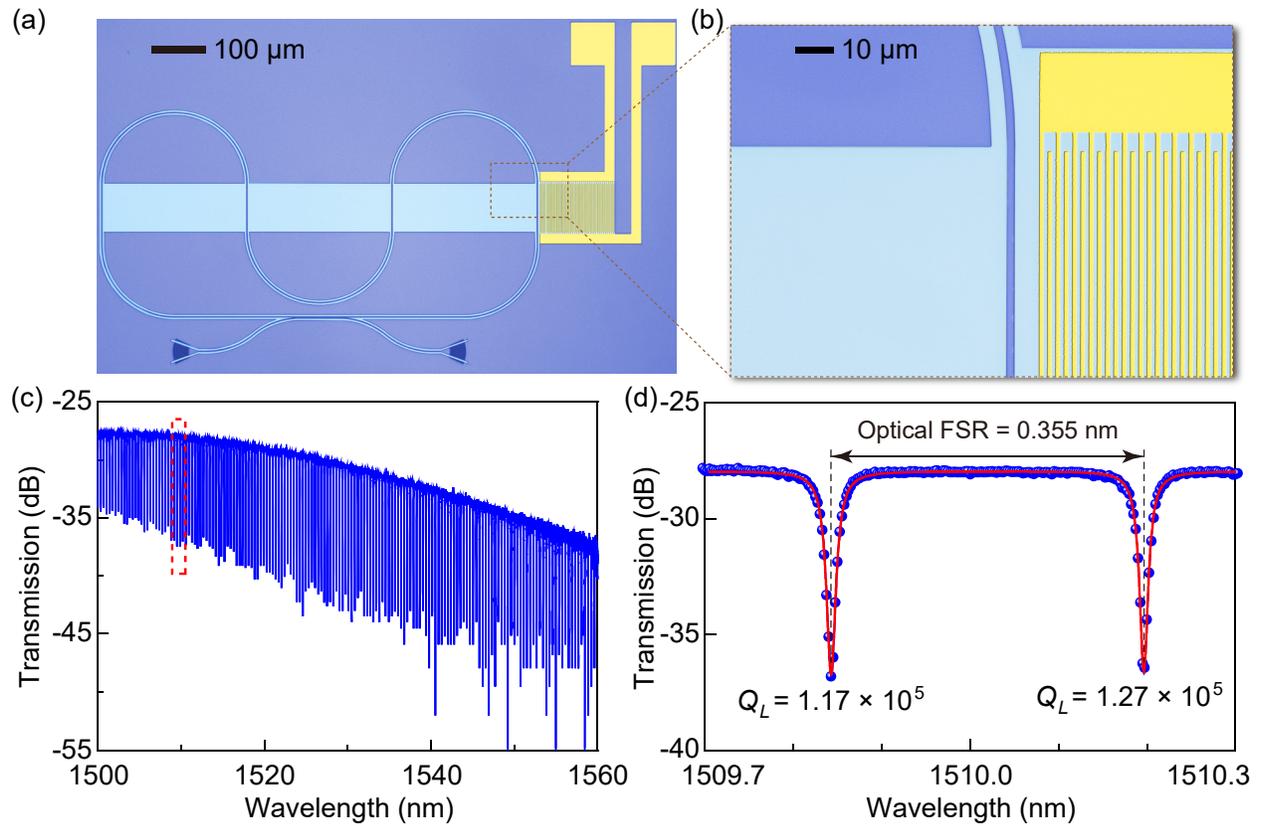

**Fig. 3.** (a) Optical microscope image of a fabricated 4-tap surface acoustic microwave photonic filter. (b) Close-up view of the acousto-optic modulation region. (c) Measured optical transmission spectrum in a broad wavelength range. (d) Zoomed-in optical transmission spectrum near 1510 nm. The blue dots represent the measured data and the red line is the corresponding Lorentzian fit.

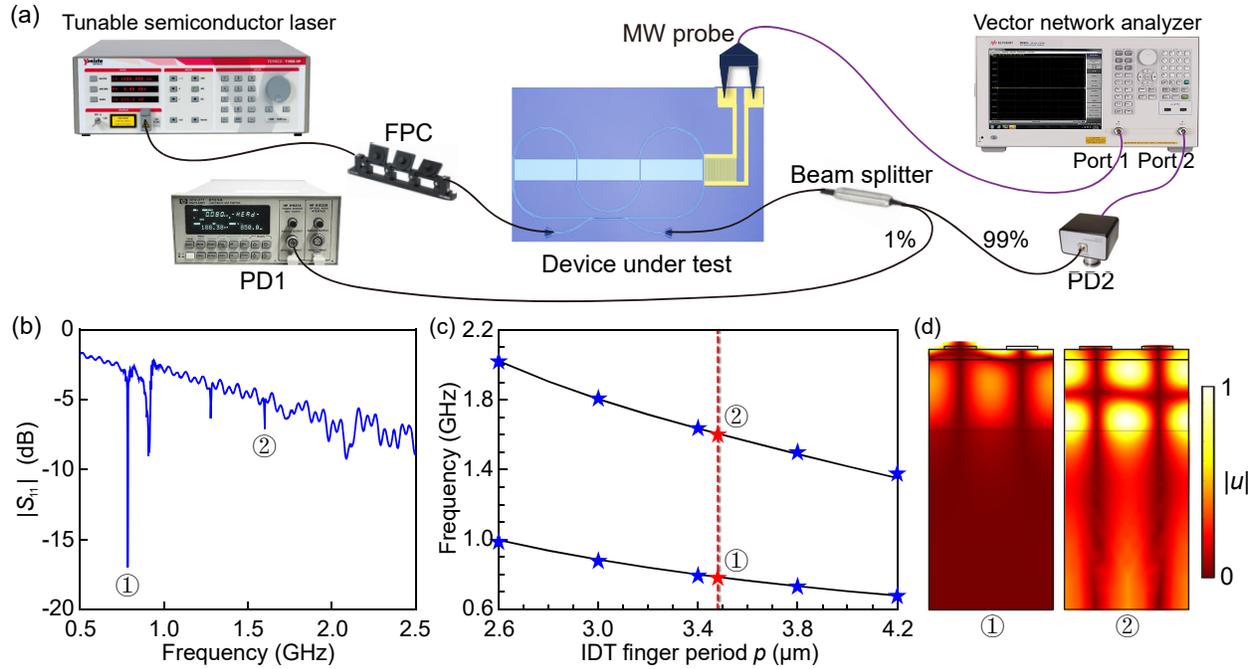

**Fig. 4.** (a) Experimental setup for measuring the acousto-optic modulation. MW probe, microwave probe; FPC, fiber polarization controller; PD1, high-sensitivity photodetector; PD2, high-speed photodetector. (b) Measured $|S_{11}|$ spectrum for a device with the IDT finger period of 3.48 μm. (c) Simulated and measured frequency of two SAWs as a function of the IDT finger period. The black solid lines plot the simulated results and the stars represent the experimentally measured data. The two red stars mark the frequencies of the two labeled SAW modes in (b). (d) Simulated cross-sectional $x$-displacement profiles $|u|$ of the two labeled SAW modes in (b).

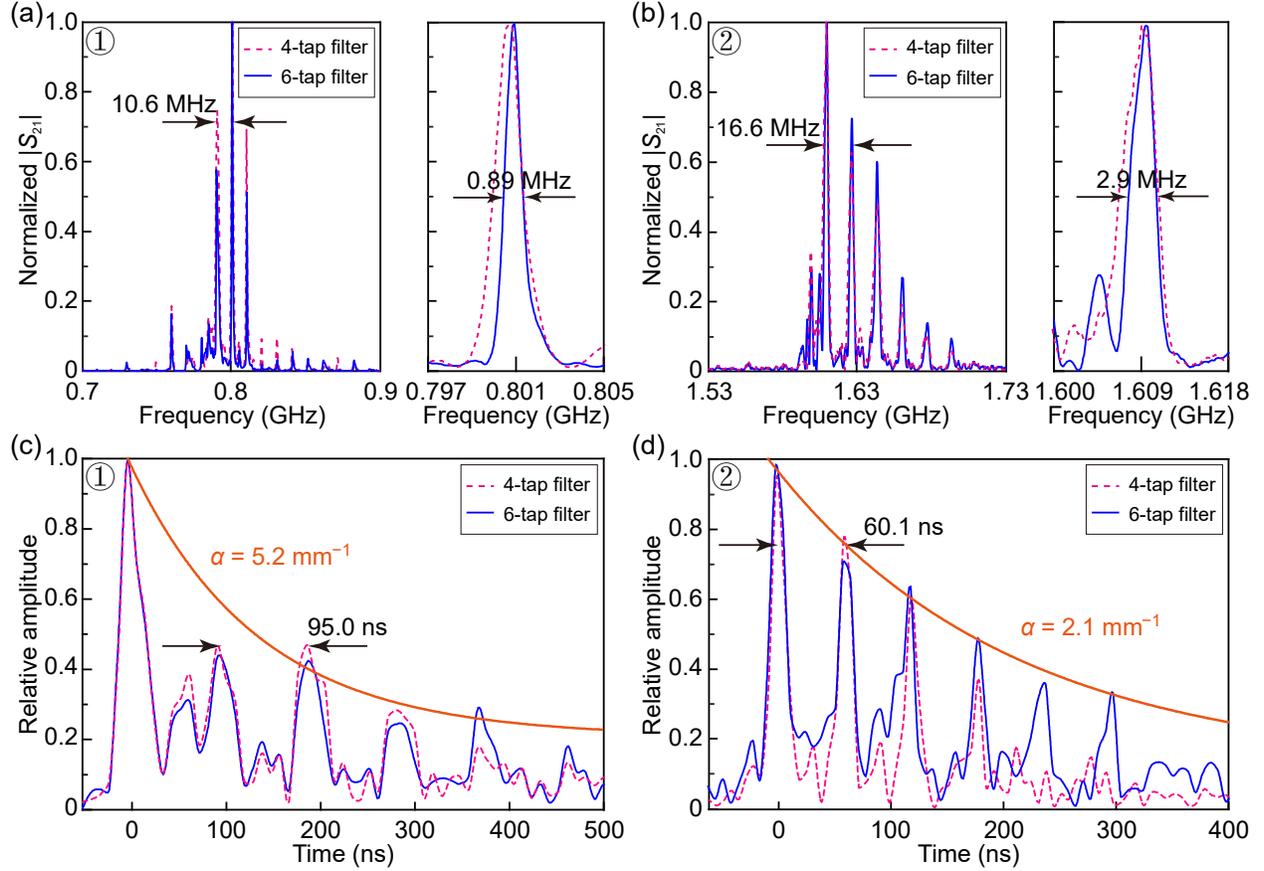

**Fig. 5.** (a),(b) Measured normalized $|S_{21}|$ spectra for 4- and 6-tap surface acoustic microwave photonic filters at frequencies near 0.78 GHz (a) and 1.60 GHz (b). The right figures show the zoomed-in spectra of the central peak, where a full width at half maximum of 0.89 MHz and 2.9 MHz was obtained in the 6-tap filter. Here both the 4- and 6-tap filters have $d = 2R = 268$ μm. (c),(d) Normalized impulse response, derived from the measured complex-valued frequency response in (a) and (b). The orange solid lines plot the corresponding exponential fits.